\documentclass[showpacs,twocolumn,nofootinbib]{revtex4} 
\usepackage{amsmath,bm,graphicx,psfrag,layout,float}
\newcommand \beq{\begin{eqnarray}}         
\newcommand \eeq{\end{eqnarray}}    

\begin{document}      
\title{Higgs Mechanism  with Type-I\hspace{-.1em}I  Nambu-Goldstone Bosons\\     
 at Finite Chemical Potential} 
\author{Yusuke Hama$^1$, Tetsuo Hatsuda$^{1,2,3}$, and Shun Uchino$^{4}$}
 
\affiliation{$^1$Department of Physics, The University of Tokyo, Tokyo 113-0033, Japan \\
 $^2$IPMU, The University of Tokyo, Chiba 277-8568, Japan\\
 $^3$Theoretical Research Division, Nishina Center, RIKEN, Saitama 351-0198, Japan\\
 $^4$Department of Physics, Kyoto University, Kyoro 606-8502, Japan }
\begin{abstract}  
  When the spontaneous symmetry breaking occurs
  for systems without Lorentz covariance, there arises  possible
   mismatch, $N_{\rm NG} < N_{\rm BG}$, 
   between numbers of Nambu-Goldstone (NG) bosons ($N_{\rm NG}$) and 
  the numbers of broken generators ($N_{\rm BG}$).
  In such a situation,  so-called  type-I\hspace{-.1em}I  NG bosons emerge. 
  We study how the gauge bosons acquire masses through the Higgs 
  mechanism under this mismatch by employing 
  gauge theories with complex scalar field at finite 
  chemical potential  and by enforcing ``charge" neutrality.
  To separate the physical spectra from unphysical ones,
  the $R_{\xi}$ gauge is adopted. Not only  massless NG bosons
   but also massive scalar bosons generated by the chemical potential
   are absorbed into spatial components of the gauge bosons. 
  Although the chemical potential induces a non-trivial mixings 
   among the scalar bosons and temporal components of the gauge bosons, 
   it does not  affect the structure of the physical spectra,
    so that  the    total number of physical modes is
   not modified even for $N_{\rm NG} < N_{\rm BG}$.
\end{abstract}
\maketitle
\section{ Introduction}

The spontaneous symmetry breaking (SSB) and the Higgs mechanism are
 the two key concepts in  both elementary particle physics and condensed matter physics.
One of the most important aspects of SSB is the appearance of massless 
 Nambu-Goldstone  (NG) bosons \cite{Nambu:1961tp,Goldstone:1961eq}: 
  In particular, for systems with 
 Lorentz covariance, the number of NG bosons, $N_{\rm NG}$, is 
 equal to the number of broken generators, $N_{\rm BG}$, of the 
 symmetry group under consideration \cite{Goldstone:1962es}. If the symmetry is local, 
 these NG bosons are absorbed into the gauge bosons and disappear from the 
  physical spectra \cite{Nambu:1960tm}.  However, for the system without Lorentz covariance,
 there arise situations with  $N_{\rm NG} \neq N_{\rm BG}$:
 A well-known example is the Heisenberg ferromagnet where
 there is only one NG magnon while the number of broken generator
  associated with O(3)$\rightarrow$O(2) is two, i.e. $N_{\rm NG} < N_{\rm BG}$ 
  (see e.g. \cite{Leutwyler:1993gf}).
 This is in contrast to the  Heisenberg antiferromagnet which shows the same symmetry 
 breaking pattern but has
  two magnons, i.e. $N_{\rm NG} = N_{\rm BG}$ (See TABLE \ref{tab:NO-NG}).
\begin{table*}[!htb]          
\label{tab:NO-NG} 
\begin{center}
\begin{tabular}{|c|c|c|c|c|c|} 
\hline
system & SSB-pattern &  $N_{\text{BG}}$ & $N_{\text{NG}}$ &NG boson & dispersion relation \\ \hline \hline
2-flavor QCD & $\text{SU(2)}_{\text{L}} \times \text{SU(2)}_{\text{R}} \rightarrow \text{SU(2)}_{\text{V}}$  & 3 & 3 & pion & $E(p) \propto p$ \\ \hline
Heisenberg antiferromagnet & $\text{O}(3) \rightarrow \text{O}(2)$ &  2 & 2 & magnon & $E(p) \propto p$ \\ \hline
Heisenberg ferromagnet & $\text{O}(3) \rightarrow \text{O}(2)$ &  2 & 1 & magnon & $E(p) \propto p^2$ \\
\hline 
\end{tabular}   
\caption{Examples of SSB. $N_{\text{NG}}$ and $N_{\text{BG}}$ denote the total number of NG bosons and broken generators, respectively.} 
\end{center}
\end{table*}  

 It was realized by  Nielsen and Chadha \cite{Nielsen:1975hm} that such a  mismatch between 
 $N_{\rm NG}$ and $N_{\rm BG}$ as the ferromagnet is related to the dispersion relation of the 
 NG bosons. By introducing type-I and type-I\hspace{-.1em}I  NG bosons
  according to whether the dispersion relation is proportional to odd and even powers
 of momentum in the long wavelengths,  they have shown an inequality,
 $N_{\rm I}+2 \times N_{\rm I\hspace{-.1em}I } \ge N_{\text{BG}}$
 where $N_{\rm I}$ ($N_{\rm I\hspace{-.1em}I }$)
 is the total numbers of type-I (type-I\hspace{-.1em}I) NG bosons.   
 The magnon in the antiferromagnet (ferromagnet) is type-I (type-I\hspace{-.1em}I),
  in this classification.  The kaon condensation in 
  the color-flavor-locked (CFL) phase of high density quantum chromodynamics  (QCD)
   shows another example
   of this mismatch. It is 
   a relativistic system with Lorentz covariance explicitly broken by 
   chemical potential and  has both  type-I and
 type-I\hspace{-.1em}I  NG bosons \cite{Miransky:2001tw}
    (See also, \cite{Blaschke:2004cs}). An important role of the commutation relations among 
     broken generators for the emergence of the type-I\hspace{-.1em}I   NG bosons was also 
     realized in this context  as reviewed in Ref. \cite{Brauner:2010wm}.  

 A natural question to ask in the presence of local gauge symmetry is the  
  fate of the gauge bosons and Higgs mechanism with type-I\hspace{-.1em}I  NG bosons.
   For  systems with  $N_{\rm NG}=N_{\rm I}=N_{\rm BG}$,
  the number of massive gauge bosons (except for the spin degrees of freedom)
  due to Higgs mechanism is  equal to the number of broken generators.
    On the other hand,
  for the systems with
  $N_{\rm NG} <  N_{\rm BG}$, it is not entirely obvious
 how the Higgs mechanism works and what would remain in the 
 physical spectra at low energies.
  In this paper, we study the Higgs mechanism with
  type-$\mbox{I\hspace{-.1em}I}$  NG bosons in relativistic systems that 
Lorentz covariance is explicitly broken by chemical potential.
 In our analysis, we employ gauge theories with complex scalar field
  at finite chemical potential such as
  gauged SU(2) model,  Glashow-Weinberg-Salam type gauged U(2) model, 
  and gauged SU(3)  model, which are known to have
   both type-I and type-I\hspace{-.1em}I  NG bosons if gauge couplings are absent.  
  To ensure the non-Abelian charge neutrality of the system, we introduce
   non-Abelian external sources according to \cite{Kapusta:1981aa}.
  Then, we  derive explicitly the mass spectra of 
   the scalar bosons and gauge bosons in the tree level.
  To separate physical spectra from unphysical ones clearly,
  we adopt  the $R_{\xi}$ gauge with the gauge  
   parameter taken  as infinity at the end.
   \footnote{
  In ref.\cite{Gusynin:2003yu}, the similar  problem was treated without
  imposing the non-Abelian charge neutrality.
   In such an approach, the  
  temporal component of the gauge field acquires non-vanishing expectation
  value in contrast to ours.  This leads to   the 
  dispersion relations of the physical modes
  and the  behavior of the system near the weak gauge-coupling limit
  different from ours (see Appendix for details).}

   This paper is organized as follows. In Sec. \ref{sec:II}, 
we briefly review NG boson spectra at finite chemical potential by taking the U(2) model
 of complex scalar fields. In Sec. \ref{sec:III}, 
 we delineate how the Higgs mechanism including the type-I\hspace{-.1em}I  NG bosons
  works through this model by gauging the  SU(2) symmetry.    
  The Glashow-Weinberg-Salam type gauged U(2) model is also studied.
In  Sec. \ref{sec:IV}, we  discuss U(3) model with its SU(3) part gauged
 as a toy model for the two-flavor color superconductivity (2SC) in dense QCD.
  In all these models,  we introduce  the chemical potential for U(1) (hyper) charge.
Section \ref{sec:V} is devoted to summary and concluding remarks.
 In Appendix, we make a detailed comparison of the results of Higgs mechanism at finite chemical 
 potential with and without the  background charge density by taking the 
  gauged SU(2) model as an example.

\section{Type-I\hspace{-.1em}I  NG Bosons in U(2) Model}
\label{sec:II}

\begin{figure}[t]
\begin{center}  
\includegraphics[scale=0.8]{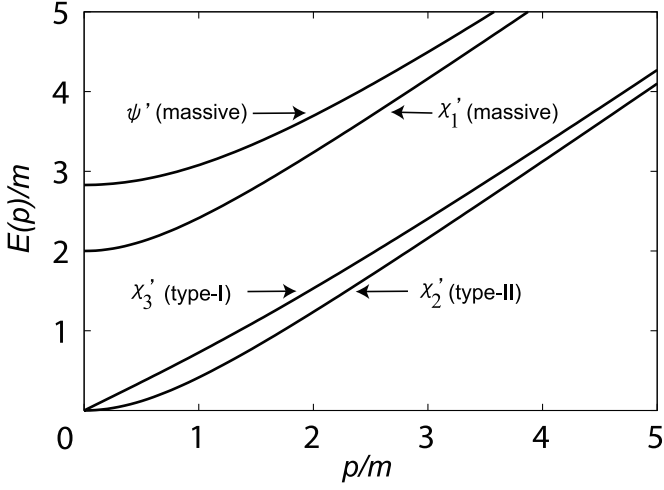}  
\caption{Dispersion relation for the fluctuation fields
 in the case of $\mu/m=1$. 
%The qualitative behavior of dispersions are same for any $\mu/m>0$. 
Due to the mixing induced by the 
 chemical potential, there arise only two NG bosons instead of three;
  one is the type-I ($\chi_3'$) with $E \propto p$
   and the other is type-I\hspace{-.1em}I    ($\chi_2'$)
    with $E \propto p^2$ at low momentum.}
\label{fig:disp}
\end{center} 
\end{figure}

Let us first review the NG boson spectra at finite chemical potential
 through the U(2) model  of complex scalar fields. It was originally introduced 
  as a model for kaon condensation in
the color-flavor-locked (CFL) phase of high density QCD \cite{Miransky:2001tw}.
The Lagrangian of the U(2)  model is given by 
\begin{eqnarray}
{\cal L} 
=|(\partial_0 -i\mu)\phi|^2 - |\partial_i  \phi|^2 
+m^2 |\phi|^2 -\lambda |\phi|^4, 
\label{eq:sec21}
\end{eqnarray} 
with $m^2$ and  $\lambda$ being positive, and 
$\phi = (\phi_1, \phi_2)^t$ denoting 2-component complex scalar field.
  Equation \eqref{eq:sec21} possesses U(2) ($\cong \text{SU}(2) \times \text{U}(1)$) symmetry, 
$\phi^{\prime}=\left[ \text{exp} (-i\theta_{\alpha}\tau^{\alpha})\right] \phi, 
\  (\alpha=0,1,2,3),$
where $\tau^{\alpha}$s are the U(2) generators. Under the SSB pattern
 U(2)$\rightarrow$U(1)$_{Q}$ ($Q=\frac{1}{2}(1+\tau_3)$) obtained by the 
 ground state expectation values,
 $\langle \phi_1 \rangle =0$ and $\langle \phi_2 \rangle =v/\sqrt{2}$
  with $v=\sqrt{{(\mu^2+m^2)}/{\lambda}}$, 
  we may parametrize the scalar field as
\begin{eqnarray}
\phi&=&\frac{1}{\sqrt{2}}\left[v+\psi+i\sum_{a=1}^{3}\chi_a \tau^a \right]\left( 
\begin{array}{c} 
0  \\
1  \\
\end{array} 
\right) ,   \nonumber\\ 
&=&\frac{1}{\sqrt{2}}\left( 
\begin{array}{c} 
\chi_2+i\chi_1  \\
v+\psi-i\chi_3  \\
\end{array} 
\right). \label{param}
\end{eqnarray} 
 Quadratic part of the Lagrangian for 
 the fluctuation fields reads 
\begin{eqnarray}
{\cal L}_0 &=& \frac{1}{2}(\partial_\mu \chi_a)^2+\frac{1}{2}\left[(\partial_\mu \psi)^2-2(\mu^2+m^2) \psi^2\right] \nonumber\\
&-&\mu(\chi_3 \overleftrightarrow{\partial_0} \psi+\chi_2 \overleftrightarrow{\partial_0} \chi_1), \label{lag}
\end{eqnarray}  
with a notation, 
$A \overleftrightarrow{\partial_0} B \equiv A \partial_0 B - (\partial_0 A) B$.
 Then the equations of motion for $\psi$ and $\chi_a$ are given by
\begin{eqnarray}
\left[ 
\begin{array}{cc}
\partial^2_0 - \partial_i^2 & -2\mu \partial_0 \\
2\mu \partial_0 &  \partial^2_0 - \partial_i^2  \\
\end{array} 
\right] \left[ 
\begin{array}{c}
{\chi}_1  \\
{\chi}_2 \\
\end{array} 
\right]
&=& 0, \\
\left[ 
\begin{array}{cc}
 \partial^2_0 - \partial_i^2 +2(\mu^2+m^2) & -2\mu \partial_0              \\
 2\mu \partial_0                           &  \partial^2_0  -\partial_i^2  \\
\end{array} 
\right] \left[ 
\begin{array}{c}
\psi  \\
{\chi}_3 \\
\end{array} 
\right]
&=& 0. 
\label{eq:sec23}
\end{eqnarray} 
Solving these equations in momentum space
 leads to the dispersion relations: 
\begin{eqnarray}
 E_{\chi_1 \text{-} \chi_2} &=& \sqrt{ p^2+\mu^2 }
\pm\mu =
\left\{ \begin{array}{l}
\frac{p^2}{2\mu}+O(p^4), \\
 2\mu+O(p^{2}).  \\
\end{array} \right.
\label{eqa:sec23} \\
E_{\chi_3 \text{-} \psi} &=& 
   \left[ p^2 + \frac{ m_{\psi}^2 }{2}  
 \pm  \frac{m_{\psi}^2 }{2} \sqrt{ 1+\frac{16\mu^2 p^{2}}{m_{\psi}^2} } 
\right]^{1/2} \nonumber\\  
&= &
\left\{ \begin{array}{l}
\sqrt{ \frac{\mu^2+m^2}{3\mu^2+m^2} }{p} + O(p^2), \\
\sqrt{ 6\mu^2+2 m^2  } + O(p^2),          \\
\end{array} \right. 
\label{eq:sec22}
\end{eqnarray}
where $m_{\psi} \equiv \sqrt{6\mu^2+2m^2}$.
These dispersion relations are shown
 in Fig. \ref{fig:disp}. From the mixing between $\psi$ and $\chi_3$
  induced by the chemical potential $\mu$, 
    one massive mode $\psi'$ and one massless mode $\chi_3'$ arise.
   The latter is the type-I NG boson 
   whose energy is proportional to $p$.
  On the other hand,   from the mixing between $\chi_1$ and $\chi_2$
   induced by $\mu$,
   one massive mode $\chi_1'$ and  one massless mode
    $\chi_2'$ arise. The latter  is the type-I\hspace{-.1em}I  NG boson  whose
 energy is proportional to $p^2$ in the low-momentum limit. 
Although we have  $ N_{\rm NG}=2$ which is smaller than 
$N_{\rm BG} =3 $,  the Nielsen-Chadha relation is satisfied as an
 equality:
\begin{eqnarray}
N_{\rm I}+2 \times N_{\rm I\hspace{-.1em}I }=1+2 
\times 1 = N_{\text{BG}}.  
\end{eqnarray} 

\section{Gauged SU(2) Model at Finite $\mu$}
\label{sec:III}
 
In this section, by gauging the  SU(2) part of the U(2) model introduced in the 
 previous section, we discuss the Higgs mechanism at finite chemical
 potential with a
type-\mbox{I\hspace{-.1em}I} NG boson.  Fate of the gauge bosons with only
  two NG bosons is of our central interest here as we mentioned in the Introduction.
The Lagrangian of the gauged SU(2)  model with finite chemical potential is given by  
\begin{eqnarray}
{\cal L}  &=&  -\frac{1}{4}\left({F}_a^{\mu \nu}\right)^2+|(D^0 -i\mu)\phi|^2 - |D^i  \phi|^2 
 \nonumber\\
& & +m^2 |\phi|^2- \lambda |\phi|^4+ gj^a_{ \mu}A_a^{\mu}, \label{lagsu2}
\end{eqnarray} 
 where   $F_a^{\mu\nu}=\partial^\mu A_a^\nu-\partial^\nu A_a^\mu
+g\epsilon_{abc}A^\mu_b A^\nu_c$ and $D^\mu  
= \partial^\mu-i\frac{g}{2}\tau^a A^\mu_a $
  with  $g$ and $\tau^a$ ($a=1,2,3$) being the
  gauge coupling and SU(2) generators, respectively. 
  $j^a_{\nu } = j^a_{0} \delta_{\nu 0} $ is a background non-Abelian charge
 density  to ensure the charge neutrality   \cite{Kapusta:1981aa}. 
%  (See Appendix for more details about this point.)

 We take  the same parametrization as Eq.~\eqref{param} for the scalar fields
 and adopt the gauge condition (the $R_{\xi}$ gauge),
\begin{eqnarray}
F_a=\frac{1}{\sqrt{\alpha}}(\partial_\mu A_a^{\mu}+M \alpha \chi_a) \ \ \ \ \ (a=1,2,3) ,
\label{gff}
\end{eqnarray}
 with $M=\frac{g}{2}v= \frac{g}{2} \sqrt{(\mu^2+m^2)/\lambda}$ 
 and $\alpha$ being the gauge parameter. 
 The chemical potential $\mu$ is embedded in $F_a$ implicitly through $M$. 
 An advantage  of taking the $R_{\xi}$ gauge is that one can 
  clearly separate  the physical and  unphysical degrees of freedom;
    masses of unphysical particles go to infinity  and 
  decouple from physical particles in
   the limit  $\alpha\rightarrow\infty$. As we see shortly, this is particularly 
  useful to analyze the situation with  new mixing terms 
   induced by the chemical potential.

 With the above gauge condition, the quadratic part of the Lagrangian 
  with the ghost fields (${c}_a$ and  $\bar{c}_a$)  reads
\begin{eqnarray}
 {\cal L} _0 &=&   - \frac{1}{4}(\partial^{\mu} A_a^{\nu}-\partial^{\nu} A_a^{\mu})^2 
  +\frac{1}{2}M^2 ({A}_a^{\mu})^2 
  -\frac{1}{2\alpha}({\partial}_{\mu} { A_a^{\mu}} )^2 \nonumber\\
&+&\frac{1}{2}[(\partial_{\mu}\psi)^2-2({\mu}^2+m^2){\psi}^2] 
    \nonumber \\   
&+& i\overline{c}_a(\partial^2+\alpha M^2)c_a+ \frac{1}{2}[(\partial_{\mu}{\chi_a})^2- \alpha M^2 \chi_a^{2}] \nonumber \\
  &-&\mu({\chi}_3 \overleftrightarrow{\partial_0} \psi +
   {\chi}_2 \overleftrightarrow{\partial_0} {\chi}_1 ) \nonumber \\
  &-&2\mu M(-{\chi}_2 A_1^{\nu=0}+{\chi}_1 A_2^{\nu=0}+\psi A_3^{\nu=0}).
\label{eq:sec34} 
\end{eqnarray} 
The first three lines in Eq.~(\ref{eq:sec34}) are the standard Lagrangian
in the $R_{\xi}$ gauge except for the implicit  $\mu$ dependence in $M$. 
 The fourth line is a mixing  induced by the 
  chemical potential which  leads to the
   type-I\hspace{-.1em}I NG boson as discussed in Sec. II.
 The fifth line is a mixing of  gauge fields with massless and
  massive scalar bosons induced by the chemical potential. 
 Note here that the linear term of the gauge field, $-\mu MvA_3^{\nu=0}$, arising from 
 $|(D_0-i\mu)\phi|^2 $ is cancelled by the  
  the background charge contribution $gj^a_{\nu }A_a^{\nu}$  with 
  $gj^{a}_{\nu}=\mu M v \delta^{a3}_{\nu 0}$.

\begin{figure*}[t] 
\begin{center}   
\includegraphics[scale=0.8]{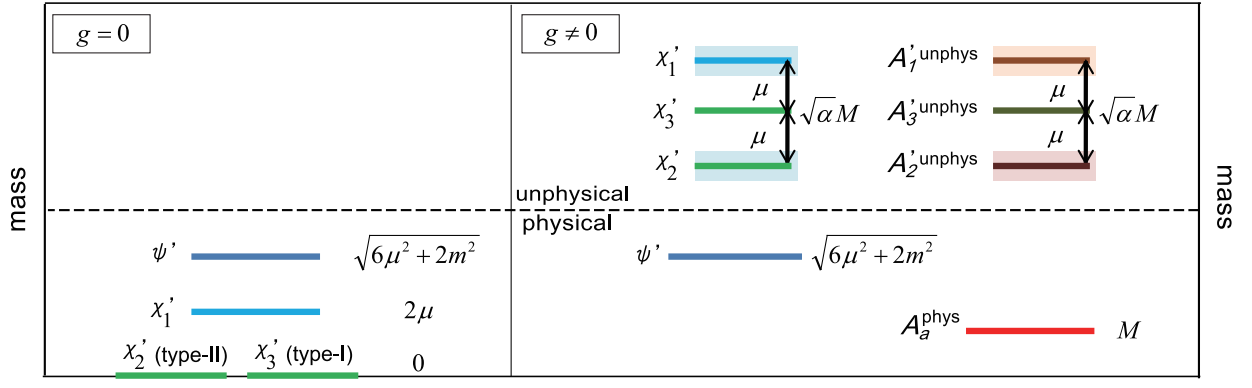}     
\caption{The left side shows the mass spectrum at $g=0$ where gauge field and the 
 scalar fields do not interact. The physical particles are 
a type-I NG boson ($\chi_3^\prime$), a type-I\hspace{-.1em}I NG boson ($\chi_2^\prime$)
 a former NG boson ($\chi_1^\prime$), and   a Higgs scalar ($\psi$). 
 In addition, there are six  (transverse$\times 3$)  massless gauge bosons.   
 The right side shows the mass spectrum at $g \neq 0$ where
  the SU(2) gauge symmetry is spontaneously broken.  
    The physical modes are the nine ((transverse+longitudinal)$\times 3$ ) 
    massive gauge bosons  $A_{a}^{\nu=1,2,3}$ with a mass $M$  
    and the Higgs scalar $\psi^\prime$. Others are unphysical modes with the 
     mass of about $\sqrt{\alpha}M$. }          
\label{fig:mass1} 
\end{center}   
\end{figure*}   

From Eq.~\eqref{eq:sec34}, 
 one finds that the spatial
 components of the gauge field $A_{a}^{\nu=1,2,3}$ absorb
   not only  massless NG bosons
   but also massive scalar boson generated by the chemical potential
  and  acquire the mass $M$.  
  On the other hand, the temporal
 component of the gauge fields $A_a^{\nu=0}$ and 
 the scalar fields ($\chi_{a}$ and $\psi$) mix with each other
  through the chemical potential. Therefore, it is important to check
  how the  mixing affects  the physical and unphysical spectra
   of the system.   For this purpose, we take $p=0$ and 
   examine the  equations of motion obtained from Eq.~(\ref{eq:sec34}), 
\begin{eqnarray}
& &\left[ 
\begin{array}{cccc}
E^2-\alpha M^2       &       -2i \mu E      &     0                  & -2\mu \sqrt{\alpha}M  \\
      2i \mu E       &       E^2-\alpha M^2 &  2\mu \sqrt{\alpha}M   &   0                   \\
             0       & -2\mu \sqrt{\alpha}M &  E^2- \alpha M^2       &   0                   \\
2\mu \sqrt{\alpha} M &        0             &     0                  &   E^2-\alpha M^2      \\
\end{array} 
\right] \nonumber\\
 & &\times \left[ 
\begin{array}{c}
{\chi}_1  \\
{\chi}_2 \\
\frac{1}{\sqrt{\alpha}}A_1^{\nu=0}  \\
\frac{1}{\sqrt{\alpha}}A_2^{\nu=0} \\
\end{array} 
\right]  
\equiv {\cal{M}}_1 \vec{X}=0, \label{eq:A12}\\ 
& &\left[ 
\begin{array}{ccc}
E^2 - 2(\mu^2+m^2)    &   -2i \mu E      &   -2\mu \sqrt{\alpha}M   \\
2i \mu E              & E^2-\alpha M^2   &       0                  \\
2 \mu \sqrt{\alpha}M  &         0        &  E^2- \alpha M^2         \\ 
\end{array} 
\right] \nonumber\\ 
& & \times  \left[ 
\begin{array}{c}
{\psi} \\
{\chi}_3 \\
\frac{1}{\sqrt{\alpha}}A_3^{\nu=0}  \\
\end{array} 
\right]  
\equiv {\cal{M}}_2 \vec{Y}=0. \label{eq:A3}
\end{eqnarray} 

 Equation (\ref{eq:A12}) implies that $\chi_{1,2}$ and $A_{a}^{\nu=0}$ have 
  a large diagonal (mass)$^2$ of $O(\alpha)$ for large $\alpha$.
  Also there are  off-diagonal terms
   of $O(\sqrt{\alpha})$, i.e. 
   $ \pm 2i \mu E \simeq \pm 2i \mu M \sqrt{\alpha}$ (for the 
   $\chi_1$-$\chi_2$ mixing)
    and 
   $ \pm 2 \mu \sqrt{\alpha} M$ (for the $\chi_{1}$-$\frac{1}{\sqrt{\alpha}} A_{2}^{\nu=0}$
    and $\chi_{2}$-$\frac{1}{\sqrt{\alpha}} A_{1}^{\nu=0}$ mixings).
  By approximating  $E$ by $\sqrt{\alpha}M$ in the off-diagonal terms
   (which is justified for large $\alpha$),  
 one can solve ${\rm det}\ {\cal{M}}_1=0$ and obtain
\begin{eqnarray}
  E_{\chi_{1,2} \text{-} A_{1,2}^0}
  =\sqrt{\alpha }M\left(1\pm \frac{2\mu}{\sqrt{\alpha}M} e^{\pm i\pi/3} \right)^{1/2}. \label{fieldmass1} 
\end{eqnarray}   
This result shows that both $\chi_{1,2}$ and $A_{1,2}^{\nu=0}$ 
 decouple from 
 physical particles due to their large masses of $O(\sqrt{\alpha})$ 
 with a small and complex mass splittings of $O(\mu)$.
  For Eq. (\ref{eq:A3}),
   ${\rm det}\ {\cal{M}}_2=0$  can be solved exactly as
\begin{eqnarray}
 E_{\chi_3 \text{-} A_3^0 } &=&
  \sqrt{\alpha} M, \label{mass1}\\
 E_{\psi'} &=&  
   \sqrt{6\mu^2+2m^2}. \label{fieldmass2}
\end{eqnarray}  
This shows that $A_3^{\nu=0}$ and $\chi_3$ decouple from physical particles due to 
their large masses  of $O(\sqrt{\alpha})$, while 
 $\psi$ remains as a physical particle with a mass not modified at all by the 
  mixing due to $\mu$.
 
 Taken together, there arise six unphysical modes:
  not only the type-I and type-I\hspace{-.1em}I  NG bosons ($\chi^\prime_{2,3}$) but also a massive mode  ($\chi^\prime_1$) become unphysical together with $A^{\nu=0}_{1,2,3}$
   due to  Higgs mechanism.
On the other hand, the physical modes are the spatial components of the gauge field 
$A_a^{\nu=1,2,3}$ (two transverse and one longitudinal) with the mass $M$, and the scalar 
mode  $\psi^\prime$ with its mass remaining invariant under the mixing induced by the 
chemical potential. The schematic illustration of the mass spectra is shown in
 FIG.~\ref{fig:mass1}. 
 Numbers of physical particles with and without the gauge coupling $g$
  are listed  in TABLE I\hspace{-.1em}I; the total physical degrees of freedom (=10)
   are  conserved regardless of the Higgs mechanism  at finite $\mu$.

%\onecolumn
\begin{table}[bht]    
\begin{center}
\begin{tabular}{|c|c|c|c|c|}   
\hline
chemical potential & \multicolumn{2}{|c|}{$\mu\neq 0$}  \\
\hline 
gauge coupling &  $g = 0$ & $g \neq 0$  \\ \hline \hline 
gauge bosons  &   2 $\times$ 3 & 3 $\times$ 3 \\ \hline
NG bosons   & 2 (type-I \& I\hspace{-.1em}I )
 & \ \ \ \ 0 \ \ \ \ \\ \hline
massive bosons   & 2 & 1 \\
\hline
\end{tabular}  
\caption{Comparison of the physical degrees of freedom at finite $\mu$ with and
 without the  gauge coupling $g$.}  
\end{center}
\end{table}  
%\twocolumn 

 Let us  briefly 
discuss the Glashow-Weinberg-Salam type gauged U(2) ($\cong \text{SU}(2) \times \text{U}(1)_Y$) 
model with the SU(2) gauge fields, $A_a^{\nu}$, and the U(1)$_Y$ gauge field, $B^{\nu}$.   Qualitative aspects of the Higgs mechanism at finite $\mu$
 in this case is the same as  that of the gauged SU(2) model.  
The mixing term induced by $\mu$ in the Glashow-Weinberg-Salam model reads 
\begin{eqnarray}
{\cal L}_{0}^{\rm mix}  &=& 
- \mu({\chi}_3 \overleftrightarrow{\partial_0} \psi + {\chi}_2 \overleftrightarrow{\partial_0} {\chi}_1 )\nonumber\\
& &- 2\mu M_W(-{\chi}_2 W_1^{\mu=0}+{\chi}_1 W_2^{\mu=0}) \nonumber \\
& & - 2\mu M_Z\psi Z^{\mu=0},  
\label{GWS} 
\end{eqnarray}
where $M_W=M=\frac{g}{2}v$, $M_Z=\frac{(g^2+g^{\prime 2})^{\frac{1}{2}}}{2}v$,
 $g^{\prime}$ denoting the U(1)$_Y$ gauge coupling, 
 $W_{1,2}^\mu=A_{1,2}^\mu$, 
 and $Z^\mu=(g^2+g^{\prime 2})^{-\frac{1}{2}}(gA^\mu_3-g^\prime B^\mu)$. 
 We impose both Abelian and non-Abelian charge neutralities
 by introducing
 the $\text{U}_Y(1)$ background charge density 
 ($g^\prime B^\mu j_\mu$) and the $\text{SU}(2)$ background charge density 
 ($g A^\mu_a j_\mu^a$). They have a role 
 to cancel out the tadpole term of gauge field, $-\mu M_Z vZ^{\nu=0}$.
Comparing these with Eq.~\eqref{eq:sec34}, the only modification   is the effect of 
  $g^{\prime}$, which leads  the $\chi_3^\prime$ mass to be $\sqrt{\alpha} M_Z$.  
 Physical mass spectra are essentially the same as
  FIG.~\ref{fig:mass1}  with an addition of massless photon field.

\section{Gauged SU(3) Model at Finite $\mu$}
\label{sec:IV}

In this section, we discuss the Higgs mechanism in the gauged SU(3)  model
 at finite $\mu$ whose Lagrangian  is given by the same form as 
Eq.~\eqref{lagsu2} with $\phi$ replaced by  a  3-component complex scalar field and $\tau^a$
replaced by the SU(3) generators.
 If we interpret   $\phi$ as  colored diquarks, the SU(3) gauge fields as gluons
  and the background source  $gj^a_{\mu}A^{\mu}_a$ as a contribution from unpaired quarks,
  this Lagrangian may be considered as 
 a toy model for the  two-flavor color superconductivity (2SC) in dense QCD \cite{Iida:2002ev}.
  This  explains the physical meaning of  the non-Abelian background charge density
  we have introduced.
  
 Let us parametrize  $\phi$ as   
\begin{eqnarray}
\phi &=&  \frac{1}{\sqrt{2}}[v+\psi+i\sum_{a=4}^{8}{\chi}_a\tau^a ]\left[ 
\begin{array}{c}
0  \\
0 \\
1 \\
\end{array} 
\right]\nonumber\\
&=&\frac{1}{\sqrt{2}}\left[ 
\begin{array}{c}
\chi_5 +i\chi_4  \\
\chi_7 +i\chi_6 \\
v+\psi -i\frac{2}{\sqrt{3}}\chi_8  \\
\end{array} 
\right]. 
\label{eq:sec42} 
\end{eqnarray}  
By adopting the $R_\xi$ gauge and choosing 
the background charge as 
   $gj_{a}^{\nu}=\frac{2}{\sqrt{3}}\mu M v \delta_{a8}^{\nu 0}$ to maintain
    neutrality, the mixing term induced by $\mu$ in 
    the quadratic part of the Lagrangian becomes
\begin{eqnarray}  
{\cal L} _0^{\rm mix} &=& 
- \mu \left( \frac{2}{\sqrt{3}}{\chi}_8 \overleftrightarrow{\partial_0} \psi +
   {\chi}_5 \overleftrightarrow{\partial_0} {\chi}_4+{\chi}_7 \overleftrightarrow{\partial_0} {\chi}_6 \right) \nonumber \\ 
  && -2\mu M(-{\chi}_5 A_4^{\nu=0}+{\chi}_4 A_5^{\nu=0}-{\chi}_7 A_6^{\nu=0}+{\chi}_6 A_7^{\nu=0} \nonumber\\ 
& & \ \ \ \ \ \ \ \ \ \ \ \  +\frac{2}{\sqrt{3}}\psi A_8^{\nu=0}). \label{su3lag}  
\end{eqnarray} 
If the gauge coupling $g$ is zero, 
  there arise
  three massive scalar bosons $\psi^\prime$ and $\chi^\prime_{4,6}$, two type-I\hspace{-.1em}I  NG bosons   $\chi^\prime_{5,7}$, and one type-I  NG boson $\chi^\prime_{8}$ 
  due to the mixing among scalar fields.
  The  mixing terms between the scalar fields and the 
  temporal component of the gauge fields   induced by the chemical potential lead to the 
 similar equation of motion as the SU(2) case at $p=0$, 
\begin{eqnarray} 
& &{\cal{M}}_1 
\left[ 
\begin{array}{c}
{\chi}_4  \\
{\chi}_5 \\
\frac{1}{\sqrt{\alpha}}A_4^{\nu=0}  \\
\frac{1}{\sqrt{\alpha}}A_5^{\nu=0} \\
\end{array} 
\right]
={\cal{M}}_1 
\left[ 
\begin{array}{c}
{\chi}_6  \\
{\chi}_7 \\
\frac{1}{\sqrt{\alpha}}A_6^{\nu=0}  \\
\frac{1}{\sqrt{\alpha}}A_7^{\nu=0} \\
\end{array} 
\right]={0}, \label{eom41}\\
& &{\cal{M}}_2 
\left[ 
\begin{array}{c}
{\psi} \\
\frac{2}{\sqrt{3}}{\chi}_8 \\
\frac{2}{\sqrt{3\alpha}}A_8^{\nu=0}  \\
\end{array} 
\right]={0}. \label{eom42}
\end{eqnarray} 
 By applying the same argument as given in Eq.~\eqref{eq:A12},
  $\chi_{4,5,6,7,8}$ and $A^{\nu=0}_{4,5,6,7,8}$ are found to decouple from 
 physical particles due to their large masses of $O(\sqrt{\alpha})$. 
On the other hand,  the spatial component of the gauge field $A_a^{\nu = 1,2,3}$
 with a mass $M=\frac{g}{2}\sqrt{(\mu^2+m^2)/\lambda}$
  and the scalar mode $\psi^\prime$ with a mass $m_{\psi}=\sqrt{6\mu^2+2m^2}$
  remain as physical particles.
 Total physical degrees of freedom in this case (=16)
   are  conserved regardless of the Higgs mechanism  at finite $\mu$.

\section{ Summary and conclusions}  
\label{sec:V} 

In this paper, we studied how the Higgs mechanism with type-I\hspace{-.1em}I  NG bosons
 works at finite chemical potential $\mu$ by imposing the 
  Abelian and non-Abelian charge neutrality.
 We adopt a relativistic U(2) model of two component scalar field
 which exhibits both type-I and  type-$\mbox{I\hspace{-.1em}I}$ NG bosons due to 
 the mixing term induced by the chemical potential.
 By gauging the SU(2) part of this model and  adopting  the $R_{\xi}$ gauge,
 we examined the physical and unphysical modes of the system.
 The result is schematically shown in FIG.~\ref{fig:mass1}:
 The type-I NG boson, the type-I\hspace{-.1em}I  NG boson, and 
  one of the  massive scalar boson which was a type-I NG boson at $\mu=0$
  are absorbed in the gauge bosons to create a gauge boson mass
   $M=\frac{g}{2}\sqrt{(\mu^2+m^2)/\lambda}$.
   The mass of the Higgs scalar was found to receive no
   effect despite  that it  mixes with the gauge boson and the
   type-I NG boson due to chemical potential. As a result,
  total physical degrees of freedom
   are  conserved regardless of the Higgs mechanism  at finite $\mu$.

 We applied  the above analysis to the
  the  gauged U(2) model (Glashow-Weinberg-Salam type model) and the gauged SU(3)
model (a toy model for the 2SC in dense QCD) at finite $\mu$.  Essential features 
 were found to be the same as those in the  gauged SU(2) model.
  Generalization to the U($N$) model with the 
 SSB pattern  U($N$)$\rightarrow$U($N-1$), which has
 one type-I NG boson and $N-1$ type-\mbox{I\hspace{-.1em}I}  NG bosons, is rather
  straightforward. 
   In this paper, we analyzed Higgs mechanism with type-\mbox{I\hspace{-.1em}I}  NG bosons 
in relativistic systems that Lorentz covariance is explicitly broken by chemical potential.
  It will be an interesting future problem
   to extend the present analysis for
    intrinsically nonrelativistic systems with type-I\hspace{-.1em}I NG bosons 
     such as Heisenberg  ferromagnet.

\section*{Acknowledgements} 
Y. H. thanks Naoki Yamamoto, Takuya Kanazawa, Shoichi Sasaki, Motoi Tachibana, and Osamu Morimatsu
 for useful discussions and comments.
This work  was supported in part by the
 Grant-in-Aid of  the Ministry  of Education, Science  and Technology,
 Sports and Culture (Nos. 20105003, 22340052).
\appendix
\section*{ Appendix} 
In this Appendix, we compare our results in Sec.~III and those
  of ref. \cite{Gusynin:2003yu}  by taking gauged SU(2) model as an example.
 As mentioned in Sec.~I, the difference between two approaches originates
  from the treatment of non-Abelian charge neutrality.

Before starting the comparison, we first review the case of the 
gauged U$(1)$ model at finite chemical potential following ref. \cite{Kapusta:1981aa}.
 The Lagrangian density is given by 
\begin{eqnarray}
{\cal L}
  &=&  -\frac{1}{4}\left({F}^{\mu \nu}\right)^2+|(D^0 -i\mu)\phi|^2 - |D^i  \phi|^2 
  \label{applag1}\nonumber\\
& & +m^2 |\phi|^2- \lambda |\phi|^4, 
\end{eqnarray} 
where $D^\mu = \partial^\mu-i\frac{g^\prime}{2} A^\mu $ with $g^\prime$ being the 
$\text{U}(1)$ gauge coupling. 
 Let us determine the ground state of the system 
 without the background charge density. (The gauge fixing condition such as 
 the $R_\xi$ gauge  does not affect the conclusion.)
 The equations of motion for scalar field $\phi$ and
gauge boson $A^\mu$ are 
\begin{eqnarray}
& &\! \! \! \!  -({\tilde D}_\nu{\tilde D}^\nu -m^2) \phi = 2\lambda (\phi^\dagger\phi) \phi,\\
& & \! \! \! \! \partial_\mu F^{\mu\nu } =
-i\frac{g^\prime}{2} \phi^\dagger \overleftrightarrow{\partial^\nu}\phi
- g^{\prime} \left( \frac{g^{\prime}}{2}A^\nu + \mu\delta^{\nu0} \right)
\phi^\dagger\phi ,
\end{eqnarray}
where ${\tilde D}^\nu = D^{\nu} - i\mu\delta^{\nu0}$.
By solving the above equations in the mean field approximation, 
and denoting the ground state expectation value of the scalar field as 
$\langle \phi \rangle = \phi_0 /\sqrt{2}={\rm const.} \neq 0$, 
 we have $\langle A^{\nu=i}\rangle =0$ and 
\begin{eqnarray}
& & \left( \frac{g^\prime}{2}\langle A^{\nu=0} \rangle +\mu \right)^2+m^2 = \lambda \phi_0^2 ,\label{eq:Abel-EOM-MFA-1} \\
& & \left( \frac{g^\prime}{2}\langle A^{\nu=0}\rangle +\mu \right) \phi_0^2 =0.
\label{eq:Abel-EOM-MFA-2}
\end{eqnarray}
Thus we obtain 
\begin{eqnarray}
\phi_0^2=\frac{m^2}{\lambda}, \ \ \  \langle A^{\nu} \rangle 
=-2\frac{\mu}{g^\prime} \delta^{\nu0}.
 \end{eqnarray}
 Expanding the fields around the minimums, $\phi=(\phi_0+\psi+i\chi)/\sqrt{2}$, 
 $A^\nu= {\cal A}^\nu + \langle A^{\nu} \rangle $, the quadratic part of the 
 Lagrangian becomes 
\begin{eqnarray}
 {\cal L}_{0} &=&  
  - \frac{1}{4}({\cal F}^{\mu\nu})^2 
  +\frac{1}{2}M_0^{\prime 2} ({\cal A}^{\mu}-M_0^{\prime -1} \partial^\mu \chi)^2\nonumber\\
&+&\frac{1}{2}[(\partial_{\mu}\psi)^2-2m^2{\psi}^2], 
  \end{eqnarray} 
with ${\cal F}_{\mu\nu}=\partial_\mu {\cal A}_\nu-\partial_\nu {\cal A}_\mu$
 and $M_0^\prime=\frac{g^\prime}{2} \phi_0$. We find that
  the chemical potential $\mu$  disappears from ${\cal L}_{0}$. 
   This unphysical situation is due to the absence of
    the background charge density $g^\prime j_\mu A^\mu$  \cite{Kapusta:1981aa}.
 
  Introducing the background charge density to Eq. \eqref{applag1},
  the mean-field equations \eqref{eq:Abel-EOM-MFA-1} and \eqref{eq:Abel-EOM-MFA-2} 
are modified as
\begin{eqnarray}
& &  \langle A^\nu \rangle =0 ,\\
& &  \mu^2+ m^2 
     = \lambda v^2 ,\\
& &  \frac{1}{2} \mu v^2 + j_{\nu =0}=0,
\label{eq:Abel-EOS-MFA-3}
\end{eqnarray}
with $v$ defined by
$\langle \phi \rangle \equiv v/\sqrt{2}$.  We thus find that
\begin{eqnarray}
v^2=\frac{\mu^2+m^2}{ \lambda}, \ \ \  \langle A^{\nu} \rangle =0,
 \ \ \ j_{\nu}=-\frac{1}{2}\mu v^2\delta_{\nu0}.
 \end{eqnarray}
 
 The quadratic part 
of the Lagrangian for the fluctuation fields $\psi$, $\chi$ and $A (={\cal A})$ reads 
\begin{eqnarray}
 {\cal L}_0  &=&   - \frac{1}{4}(F^{\mu\nu})^2 
  +\frac{1}{2}M^{\prime 2} ({A}^{\mu}-M^{\prime -1}\partial^\mu\chi)^2  \nonumber\\
&+&\frac{1}{2}[(\partial_{\mu}\psi)^2-2({\mu}^2+m^2){\psi}^2] 
    \nonumber \\   
  &+&\mu({\chi}_3 \overleftrightarrow{\partial_0} \psi)+2\mu M^\prime \psi {A}^{\nu=0}
\nonumber\\
&+&\mu M^\prime v{A}^{\nu=0}+g^\prime j_\mu {A}^\mu,
\end{eqnarray} 
with $M^\prime=\frac{g^\prime}{2} v$.
The total charge density of the system is the sum of 
 condensation charge and the background charge which cancel with each other:
\begin{eqnarray}
\rho_{\text{tot}}&=& \langle  \frac{\partial {\cal L}_0}{\partial {A}^{\nu =0}} \rangle \nonumber \\
&=& g^{\prime} \left( \frac{1}{2} \mu v^2+ j_{\nu=0} \right) = 0.
\end{eqnarray} 
The masses of the fluctuation fields \eqref{mass1} 
and \eqref{fieldmass2}  are obtained from ${\cal L}_0$ 
by employing the $R_\xi$ gauge.  We note that these masses approach smoothly
 to those in \eqref{eq:sec22} by taking the limit, $g^\prime\rightarrow0$.  

Now let us generalize the above discussion to the 
 gauged SU$(2)$ model.
 We first study the model without the  
background charge density following \cite{Gusynin:2003yu}. 
In this case, the Lagrangian is given by 
\eqref{lagsu2} without the term $gj_\mu^a A^\mu_a$. 
Then the equations of motions for scalar field $\phi$ and gauge bosons $A^\mu_a$ become  
\begin{eqnarray}
 & & \! \! \! \! \! \! \! \!  
 - ({\tilde D}_\nu {\tilde D}^\nu -m^2) \phi =2\lambda  (\phi^\dagger\phi )\phi,\\
& & \! \! \! \! \! \! \! \!  ({\cal D}_\mu F^{\mu\nu})^a 
= -ig \phi^\dagger \frac{\tau^a}{2}\overleftrightarrow{\partial^\nu}   \phi
 - g \phi^{\dagger} \left( \frac{g}{2}A^\nu_a + \mu\delta_{\nu0}\tau^a \right) \phi 
\nonumber\\.
\end{eqnarray}
Solving the above equations in  the mean field approximation
 with $\langle \phi \rangle =(0,\phi_0/\sqrt{2})$, we have
\begin{eqnarray}
& & \langle A^{\nu=0}_{\pm} \rangle =0, \\
& & \left( \frac{g}{2}\langle A^{\nu=0}_3 \rangle  -\mu \right)^2+m^2 = \lambda \phi_0^2, \\
& & \left( \frac{g}{2}\langle A^{\nu=0}_3 \rangle -\mu \right) \phi_0^2=0,
\end{eqnarray}
with $A^\mu_{\mp}=(A^\mu_1\pm iA^\mu_2)/\sqrt{2}$.
Then the ground states are characterized by the condensations,
\begin{eqnarray}
\phi_0^2= \frac{m^2}{\lambda}, \ \  \langle A_a^\nu \rangle 
 =2\frac{\mu}{g} \delta_{a3}^{\nu0}.
\label{eq:EP}
\end{eqnarray}
The quadratic part of Lagrangian for the fluctuation fields becomes
\begin{eqnarray}
 {\cal L}_0 &=&   - \frac{1}{4}({\cal F}^{\mu\nu}_a)^2 
  +\frac{1}{2}M_0^2 ({A}_a^{\mu}-M_0^{-1}\partial^\mu \chi_a)^2  \nonumber\\
&-&2\mu({\cal F}^2_{0i}{\cal A}_1^{i}-{\cal F}^1_{0i}{\cal A}_2^{i})
+2\mu^2\left(({\cal A}^{i}_1)^2+({\cal A}^{i}_2)^2\right)
\nonumber\\
&+&\frac{1}{2}[(\partial_{\mu}\psi)^2-2m^2{\psi}^2] 
    \nonumber \\   
  &-&2\mu({\chi}_2 \overleftrightarrow{\partial_0} \chi_1) 
+2\mu^2(\chi_1^2+\chi^2_2) \nonumber \\
  &-&2\mu M(-{\chi}_2 {\cal A}_1^{\nu=0}+{\chi}_1 {\cal A}_2^{\nu=0}),
\end{eqnarray} 
with ${\cal F}^a_{\mu\nu}=\partial_\mu {\cal A}_\nu^a-\partial_\nu {\cal A}_\mu^a$ 
and $M_0=\frac{g}{2}\phi_0 $.

Even without the background charge density, 
the SU(2) charge neutrality is still ensured by the gauge bosons having nonzero
expectation value, and we have $\rho_{\rm tot}^a 
=\langle \frac{\partial {\cal L} _0}{\partial {A}_a^{\nu=0}}\rangle =0$.
 Furthermore,  
dispersion relations for  $A^{\mu=i}_{\pm,3}$ and $\psi$ become 
\begin{eqnarray}
& &E_{A^{\mu=i}_{\mp}}^2= \left( \sqrt{p^2+(g\phi_0/2)^2} \pm 2\mu \right)^2, 
\label{appsu2dis1}\\
& &E_{A^{\mu=i}_{3}}^2=p^2+(g\phi_0/2)^2,
 \label{appsu2dis2}\\
& &E_{\psi }^2=p^2+2m^2. \label{appsu2dis3}
\end{eqnarray} 

The magnitude of the condensate of the gauge field
 in eq. \eqref{eq:EP} grows as  $g$ becomes small, so that 
 the phase characterized by eq. \eqref{eq:EP} is
  distinct from the ground state of the non-gauged U(2) model in Sec. II.
 Accordingly, the dispersion relation for the Higgs boson in \eqref{appsu2dis3} 
does not appoach to eq. \eqref{eq:sec22} in the limit $g\rightarrow0$, and 
 the gauge bosons $A^{\mu=i}_{\mp}$ are not massless in the limit $g=0$.

  We now turn to the ground state
  of the system  with the addition of SU(2) background charge density, $gj_\mu^a A^\mu_a$,
   as discussed in Sec. in Sec. III.
 By solving the Lagrangian given by \eqref{lagsu2}  
 in  the mean field approximation with $\langle \phi \rangle =(0,v/\sqrt{2})$, we obtain
\begin{eqnarray}
& &  \langle A_a^\nu \rangle =0 ,\\
& &  \mu^2+ m^2 = \lambda v^2,\\
& &  -\frac{1}{2} \mu v^2 + j_{\nu =0}^3=0.
\label{eq:Abel-EOS-MFA-5}
\end{eqnarray} 
Thus we obtain
\begin{eqnarray}
v^2=\frac{\mu^2+m^2}{ \lambda}, \ \ \  \langle A^{\nu} \rangle =0,
 \ \ \ j^a_{\nu}=\frac{1}{2}\mu v^2\delta^{a3}_{\nu0}.
\label{eq:NEP}
 \end{eqnarray} 
 
 The quadratic part of the Lagrangian for 
the fluctuation fields reads 
\begin{eqnarray}
 {\cal L}_0 &=&   - \frac{1}{4}({F}_{\mu\nu}^a)^2 
  +\frac{1}{2}M^2 ({A}_a^{\mu}-M^{-1}\partial^\mu \chi_a)^2  \nonumber\\
&+&\frac{1}{2}[(\partial_{\mu}\psi)^2-2({\mu}^2+m^2){\psi}^2] 
    -\mu({\chi}_3 \overleftrightarrow{\partial_0} \psi +
   {\chi}_2 \overleftrightarrow{\partial_0} {\chi}_1 ) \nonumber \\
    &-&2\mu M(-{\chi}_2 {A}_1^{\nu=0}+{\chi}_1 {A}_2^{\nu=0}
    +\psi {A}_3^{\nu=0})\nonumber\\
 &-&\mu Mv{A}^{\nu=0}_3+gj^a_\nu {A}^\nu_a.
\end{eqnarray} 
In this case, the SU(2) charge neutrality is ensured by the  
cancellation between the condensation charge and the background charge:
\begin{eqnarray}
 \rho^{a=3}_{\text{tot}}&=&
 \langle \frac{\partial {\cal L} _0}{\partial {A}_3^{\nu=0}} \rangle
 = g \left[ -\frac{1}{2} \mu v^2 + j^{a=3}_{\nu=0} \right] = 0, \\
 \rho^{a=1,2}_{\text{tot}}&=& 0.
\end{eqnarray}
 Adopting the $R_\xi$ gauge, and solving the equations of motions at $p=0$, 
we obtain Eqs. \eqref{eq:A12} and \eqref{eq:A3}. These equations in 
the limit $g\rightarrow0$ reproduce the masses of the scalar fields 
\eqref{eqa:sec23} and \eqref{eq:sec22}. Therefore the phase 
characterized by Eq. \eqref{eq:NEP} is smoothly connected to 
 the ground state of the non-gauged U(2) model in Sec. II.

\end{document}